# Chemical Interface Damping by Electrochemical Gold Oxidation

Maurice Pfeiffer[a]*, Xinyan Wu[a], Fatemeh Ebrahimi[ab], Nadiia Mameka[c], Manfred Eich[ab], Alexander Petrov[ab]

[a] Institute of Optical and Electronic Materials, Hamburg University of Technology, Eißendorfer Straße 38, 21073 Hamburg, Germany

[b] Institute of Functional Materials for Sustainability, Helmholtz-Zentrum Hereon, Max-Planck Straße 1, 21502 Geesthacht, Germany

[c] Institute of Materials Mechanics, Helmholtz-Zentrum Hereon, Max-Planck Straße 1, 21502 Geesthacht, Germany

* E-Mail: maurice.pfeiffer@tuhh.de, Phone: +49-40-42878-3854

**ABSTRACT**

Chemical interface damping is a change in the effective collision frequency of conduction band electrons in metal originating from a chemical change of the metal interface. In this work, we present in-situ ellipsometric measurements that reveal the chemical interface damping effect from electrochemical oxidation of single crystal and polycrystalline gold films. We observe an increase in collision frequency of up to 21 meV for single-crystalline gold. To compare to results obtained with thiols and metal-oxides on gold nanoparticles, we normalize the collision frequency by the electron mean free path to the surface of the structure. We show that electrochemical gold oxidation provides a stronger effect on collision frequency than these coatings. Similar ellipsometric experiments have previously been conducted to investigate the optical properties of gold oxide, but without taking chemical interface damping into account. The change in reflection from oxidation of gold was solely attributed to the oxide coating. We also show that the chemical interface damping effect saturates at a larger effective oxide thickness, which is attributed to the stabilization of the gold-oxide interface.

**INTRODUCTION**

In the Drude region of metal dispersion, where no interband transitions occur, intraband transitions of free electrons dominate the optical response. The probability of such transitions is described by the electron collision frequency (damping term) of the Drude model.[1] These collisions are required to compensate the change of electron momentum in the intraband transitions. In the bulk of the metal, collisions of electrons with lattice defects, phonons, and other electrons via Umklapp process contribute to the collision frequency. In nanostructures smaller than the electron mean free path, collisions of electrons with the surface provide an additional damping mechanism also called Landau damping.[2,3] Chemical interface damping (CID) is the increase in surface collision frequency due to change of the material at the interface of the metal.[4,5] CID increases the effective collision frequency by two possible mechanisms: chemistry induced effective roughness of the surface and/or direct charge transfer.[6–8]

The CID effect was first introduced by Hövel et al.[4] and has since been investigated mostly on nanoparticles covered with thiols[9–14] and metal-oxides[5,15] or in electrochemical environment[16–18]. Its scope for applications is seen to enhance photochemical processes by providing an additional pathway for plasmon decay with electrons/holes being directly transferred into hybrid states formed by the metal and the adsorbate or into a covering semiconductor,[6,19–21] before such carriers are further transferred into a surrounding electrolyte and can contribute to the external photoelectrochemical current. CID is especially interesting for increasing the efficiency of light induced chemical reactions such as water-splitting. Here, a possible application for the CID effect in nanoporous gold was discussed by Graf et al.[22] We should note that also the induced roughness mechanism can lead to the enhancement of charge transfer. Hot electrons excited close to the surface instead of the bulk have a higher probability of exiting the material and participating in chemical reactions.[2] The CID effect also offers the opportunity to create tunable metamaterials based on structures with large surface-to-volume ratios, where the contribution of CID to the overall collision frequency is significant. One such candidate is the nanoporous gold structure.[22–27]

Furthermore, CID should play a crucial role in studying electrochemically produced oxide layers on noble metals like gold. Previous investigations neglected the impact of CID and thus found the gold oxide layer to have metallic properties to explain observed absorption.[28–32] In this work, we show that the observed change in the reflection can be attributed to the increasing collision frequency of the underlying metal caused by CID.

We present an in-situ study of the optical properties of single-crystalline Au(111) with spectroscopic ellipsometry in an electrochemical environment by successfully measuring CID from an electrochemically grown oxide layer on the gold surface. The applied model consists of the gold substrate with fixed plasma frequency and variable collision frequency as well as a gold oxide layer with fixed permittivity and variable effective thickness as defined below in the Methods section. We observe a linear relation between the effective thickness of the oxide layer and the additional collision frequency of the gold up to approximately 0.8 nm. Additional growth of oxide does not result in further increase in collision frequency, indicating that the interfacial layer is complete and remains unchanged. This finding supports our approach of modelling electrochemical gold oxidation with CID. We also find the CID effect to be entirely reversible for effective oxide thickness below 0.8 nm. It should be mentioned that in ellipsometry we determine the optical thickness of the grown dielectric layer which depends on the refractive index and geometrical thickness of the layer. Here, we report the effective thickness of the layer assuming a spatially constant refractive index of 2.75. Though different anions chemisorb to the surface and different types of hydroxides and oxides are formed at different potentials[33] with different and mostly unknown optical properties, we model them all by an effective thickness with a constant refractive index. We believe this approach is reasonable, since our investigation intends to clarify the net effects of the electrochemical build-up and removal of a gold oxide layer on a single-crystalline Au(111) surface and, at this time, does not intend to resolve individual contributions of different species of gold hydroxides and oxides as well as spatial inhomogeneities in such layers.

Employing the mean free path for surface collisions of electrons, we compare our CID results obtained from planar gold to those obtained on nanoparticles. Our findings indicate that the CID effect from electrochemical oxidation is more pronounced than the one from other metal-oxide coatings and from covalently attaching thiols. An open question, which cannot be answered by this work due to the limited spectral range of our investigation, is the mechanism behind the observed CID. It is unclear which part of CID originates from an induced effective roughness of the surface and which from temporal electron transfer into oxide states. For the temporal electron transfer mechanism, one expects a certain energy barrier depending on the position of oxide levels with respect to Fermi energy of the metal.[6] Thus, the effect should disappear above certain wavelength, while the roughness mechanism is expected to be independent of photon energy. In follow-up studies, we aim to investigate a broader spectral range to find a possible energy barrier.

**METHODS**

**Gold Samples.** We utilized single crystal template-stripped (from silicon wafer) Au(111) nanolayers on glass chips purchased from Platypus Technologies. The Au layers, with thickness of 100 nm, exhibit an ultra-flat surface with minimal roughness (RMS roughness of 0.36 nm) measured by the manufacturer via scanning tunneling microscopy. Even though they are thin films they show primarily Au(111) structure (92% of grains are Au(111)) measured by the manufacturer via electron diffraction experiment. Thus, we will refer to these samples as 'single-crystalline'. The glass chips with the Au layers are carefully stripped from the silicon wafer substrate immediately prior to conducting the experiments. Additionally, we examined 100 nm thick sputtered Au films on silicon substrates with Ti adhesion layer to demonstrate similar behavior for polycrystalline films. The results from the polycrystal are given in the supporting information S9.

**Electrochemical Oxidation.** The electrochemical oxidation process of the gold surface is controlled by a potentiostat (Gamry Interface 1010E) utilizing cyclic voltammetry (CV). This method involves continuously scanning the potential between a prior set upper and lower limit. The current is simultaneously recorded and plotted against the potential, resulting in a cyclic voltammogram, which shows the oxidation behavior as positive currents and the reduction behavior as negative currents.[34] The electrochemical cell is equipped with a three-electrode setup including a pseudo-Ag/AgCl reference electrode, measured at 0.52 V versus reversible hydrogen electrode (RHE) in 0.5M $H_2SO_4$ (0.3 pH), a platinum wire counter electrode, and the gold sample connected to a thin gold wire by mechanical clamping as the working electrode. All potentials in this study are referenced to the RHE reference scale. The 0.5M $H_2SO_4$ electrolyte was degassed before running electrochemical experiments to remove resolved oxygen. The measured current and thus also the accumulated charge comprises of the net charge $Q$ that is related to the oxidation process and a charge that originates from a background current $I_b$. We assume this current to be governed by a constant resistance $R$: $I_b = \frac{U(t)}{R}$. With this we obtain the correction charge $Q_c(t) = \int_{t_0}^{t} \frac{U(t)}{R} dt = \int_{t_0}^{t} \frac{U_0 + \Delta U \cdot t}{R} dt$ with the starting potential $U_0$ and the scan rate $\Delta U$. The net charge $Q$ is then obtained from measured charge $Q_m$ as: $Q(t) = Q_m(t) - Q_c(t)$. The constant resistance is obtained based on the condition that the net charge $Q$ related to oxidation returns to zero following a complete cycle. The graphical representation of the measured and correction charge is shown in the supporting information S2.

**Ellipsometry measurements.** The optical properties of the gold samples were measured by a spectroscopic ellipsometer (SEMILAB SE-2000) equipped with an electrochemical cell for in-situ measurements. Below 1.8 eV (688 nm) no interband transitions contribute to the optical properties of gold.[35] To make sure these interband transitions do not influence our results we look at the spectral range above 750 nm, where only intraband transitions, that are described by the Drude model, influence the optical properties of gold.[35] The upper spectral limit is set by the range of the used detector to 950 nm. To ensure a balance between noise reduction and repetition speed, timed measurements were performed at five-second intervals with a two-second integration time. As we use high intensity to ensure the best possible signal to noise ratio with our setup, we exclude two spectral regions (815 nm to 830 nm and 875 nm to 890 nm) from the fitting where the detector is driven into saturation by the strong peaks of the Xe lamp used in the ellipsometer. All measurements were conducted at an angle of 68.7°.

**Optical model.** In the spectral region above 750 nm the Drude model is used to describe the optical properties of gold.[1] It is given by $\varepsilon(\omega) = \varepsilon_\infty - \omega_p^2/(\omega^2 + i\Gamma\omega)$, with plasma frequency $\omega_p$, collision frequency $\Gamma$, and the permittivity limit at high frequencies $\varepsilon_\infty$. We provide the collision frequency $\Gamma$ in energy units (electronvolt) employing $\Gamma[eV] = \hbar[eVs] \cdot \Gamma[1/s]$ with the reduced Planck constant $\hbar$. For the gold oxide layer, literature provides vastly different optical properties.[28–32,36] We use a constant refractive index of $n = 2.75$ from findings of Liu et al.[36], but omit the extinction coefficient of $k = 0.25$ for the analysis shown in the main text. The fitting results assuming lossy dielectric gold oxide are discussed in the supporting information S2. Liu et al. used reactive sputtering to produce gold oxide and determined the optical properties based on spectroscopic measurements. We prefer to use their findings compared to the ones based on electrochemistry as the former were not influenced by CID. For the ambient material of 0.5M $H_2SO_4$ electrolyte we employ the optical properties of water.[37]

**Fitting of ellipsometry data.** A spectroscopic ellipsometer measures the ratio between the reflection of s- and p-polarized light over a certain bandwidth. Fitting procedures based on the Transfer-Matrix method and the Fresnel equations are used to find a material stack, for which the theoretical result matches best with the measured one.[38] The ellipsometric results are fitted with a semi-infinite gold substrate (penetration depth much smaller than layer thickness) and a thin dielectric layer of constant refractive index and variable effective thickness $d$ on top. In the ellipsometric measurement we cannot separately obtain refractive index and thickness of the nanometric dielectric layer. We obtain the phase shift that corresponds to the accumulated optical path length in the film $\Lambda = \int_0^{x_{max}} n(x)\, dx$, where $x$ is the direction normal to the film. We define the effective geometrical thickness $d$, further called 'effective thickness' in the text, as $d = \Lambda/n$, assuming a spatially constant effective refractive index $n = 2.75$. This effective thickness considers the integral value over different types of adsorbates, e.g., hydroxide and oxide layers. The initial values for the gold substrate ($\omega_p$, $\Gamma$ and $\varepsilon_\infty$) are fitted based on the first measurements of each experiment at low potentials where no oxide is present at the surface ($d =$ 0). For the following results, only the collision frequency of gold $\Gamma$ and the oxide thickness $d$ are fitted. Additional plots regarding the fitting of the ellipsometer results with our model are given in the supporting information S1. There we show a comparison of measured and modelled ellipsometry angles Psi and Delta, RMSE values indicating the validity of the fits and a theoretical depiction on how the change of collision frequency and effective oxide thickness influence the ellipsometry angles Psi and Delta.

## RESULTS AND DISCUSSION

The measurements were carried out by running multiple subsequent CV scans while simultaneously measuring the optical properties of the gold sample with a spectroscopic ellipsometer. Overall, three subsequent electrochemical experiments were performed, all of which incorporated simultaneous in-situ measurements with the ellipsometer. The first experiment consisted of 11 CV scans with a 10 mV/s scan rate between 0.8 V and 1.7 V vs. RHE. This was coupled with ellipsometry measurements every 5 seconds (50 mV steps). The results derived from this experiment can be found in the supporting information S3. The second experiment involved CV at a slower scan rate of 1 mV/s to facilitate a more detailed study. Here, only two scans were recorded, while ellipsometry measurements were conducted every 5 seconds (5 mV steps). The same experiments have been conducted with polycrystalline sputtered gold samples. Their results are shown in the supporting information S9. The third experiment saw CV in a wider potential range, extending the upper vertex potential up to 2 V (1 mV/s with 5 mV steps) in order to generate thicker oxide layers within the regime of the oxygen evolution reaction. This unveiled a region where the collision frequency reached a plateau with increasing oxide thickness.

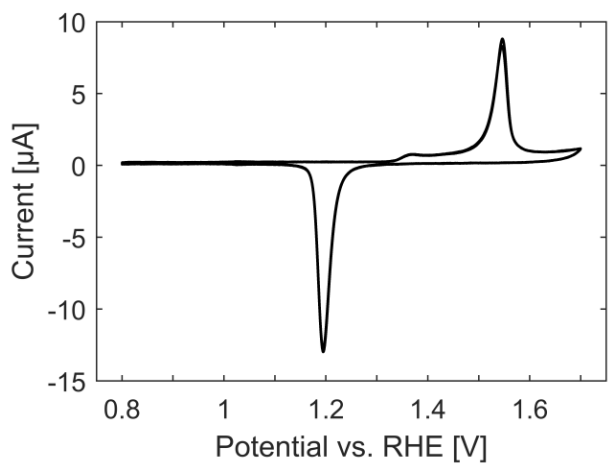

**Figure 1.** Electrochemical behavior of single crystalline gold sample. Cyclic voltammogram of template stripped Au(111) in 0.5M $H_2SO_4$ electrolyte with a scan rate of 1 mV/s showing the overlap of two subsequent scans.

Figure 1 shows the cyclic voltammogram from the second experiment with two subsequent scans. The overlap of both curves indicates the repeatability of the oxidation and reduction process. The oxidation and reduction current peaks were observed at 1.55 V during anodic scan (positive scan) and 1.18 V during the cathodic scan (negative scan). These, along with the small peak at 1.37 V, resemble the electrochemical characteristics of Au(111) in diluted $H_2SO_4$.[33,39]

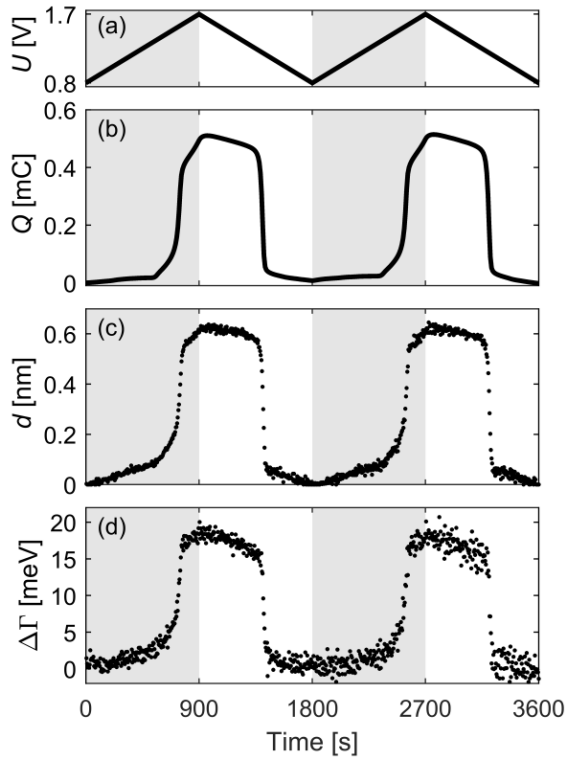

**Figure 2.** Optical measurement results compared to oxidation related charge. Reversible variations/change in (a) potential $U$, (b) net charge $Q$, (c) effective oxide thickness $d$ and (d) collision frequency change of template stripped Au(111) $\Delta\Gamma$ during two CV scans at a scan rate of 1 mV/s in 0.5M $H_2SO_4$. Subfigure (a) shows the linear scanning of the potential between 0.8 V and 1.7 V and subfigure (b) shows the net charge corrected by a background current as described in the methods section. The shaded area indicates the anodic scan region (0.8 to 1.7 V) and the white one the cathodic scan region (1.7 to 0.8 V). Plots (c) and (d) show the fitting results from the ellipsometry measurements, displaying the variation in the effective oxide thickness and the change in the collision frequency, respectively. They show a well reversible oxidation process along with a reversible change of the optical properties of the Au substrate from CID. The electrochemical oxidation results in a 0.6 nm thick oxide layer and an increase in collision frequency of 18 meV. A figure showing the reversibility over 11 CV scans with a scan rate of 10 mV/s is shown in the supporting information S3.

The current shown in Figure 1 comprises of the current associated with the oxidation and reduction reactions, along with a background current. To identify the net charge $Q$ contributing to the oxidation and reduction reactions on the gold surface, we subtract the correction charge $Q_c$ from the measured charge $Q_m$ as detailed in the methods section. A figure detailing this process is shown in the supporting information S2. $Q$ is plotted in Figure 2 together with the fitting results from ellipsometry. The fitting of the ellipsometric measurement results only uses two free parameters: the collision frequency $\Gamma$ and the effective oxide thickness $d$. We assume that the oxidation does not change the free electron density and their effective mass inside the metal, which both are linked to the plasma frequency $\omega_p$. Thus, we keep $\omega_p$ constant. A redshift of the plasmon resonance of oxidized particles was previously attributed to change of electron density and therefore $\omega_p$.[16] But this shift can be also explained by the optical effect of the oxide dielectric coating.[15,40] As the oxidation is not expected to change the gold behavior at large energies, $\varepsilon_\infty$ is kept constant as well. $\omega_p$ and $\varepsilon_\infty$ are obtained by fitting only the first measurements at low potential with no oxide layer on the gold surface. For the optical properties of the gold oxide layer, we assume a purely dielectric material as detailed in the methods section.

Figure 2 shows the correlation between the corrected net charge $Q$, the effective oxide layer thickness $d$, and the change in collision frequency $\Delta\Gamma$. We identify a linear relationship between these parameters (detailed in the supporting information S4). The increase in collision frequency originates from CID at the interface between the electrochemically produced oxide layer and the gold surface. In this experiment, the oxide layer reaches a maximum effective thickness of 0.6 nm, while the increase in collision frequency reaches 18 meV. Similar effects were observed on sputtered polycrystalline film (detailed in the supporting information S9). The obtained thickness matches well with a similar study by Kolb and McIntyre[29] as well as Horkans et al.[31] A comparable linear relationship between coverage and change in collision frequency was reported by Foerster et al.[11] for thiols. Additional plots of the fitted parameters vs. charge and vs. applied potential are shown in the supporting information S5.

Furthermore, we observe a minor decline in collision frequency and oxide thickness as soon as the cathodic scan commences. We attribute this to a slight reduction of oxide as soon as the potential drops. During the reduction phase, we then observe a steep decline in both oxide thickness and collision frequency, which return both to their initial values. This confirms the reversibility of the process.

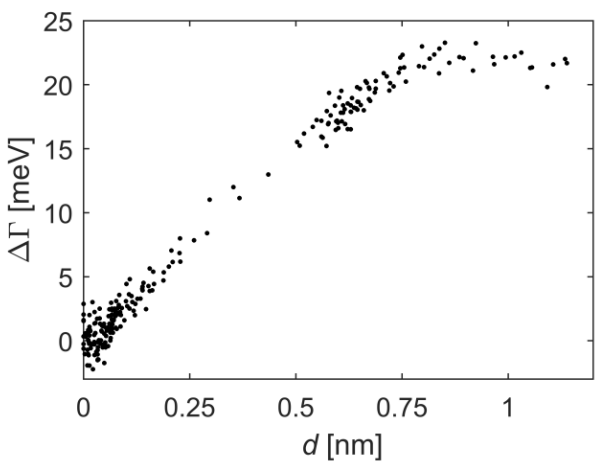

**Figure 3.** Saturation of collision frequency at larger effective oxide thickness. Change of collision frequency $\Delta\Gamma$ of template stripped Au(111) vs. effective oxide thickness $d$ during a CV scan between 0.8 and 2 V (scan rate: 1 mV/s) in 0.5M $H_2SO_4$, going into the regime of oxygen evolution reaction. The corresponding cyclic voltammogram is shown in the supporting information S6. We observe a linear relation between the collision frequency and the oxide

thickness up to 0.8 nm followed by saturation. This result matches well with our model for Au oxidation including the CID effect.

For the effective oxide thicknesses obtained in the experiment shown in Figure 2, we observe a linear relationship with the collision frequency. To obtain thicker oxide and confirm our model of collision frequency saturation, we also apply higher potentials. Figure 3 shows the relationship between the oxide thickness and the collision frequency during an anodic potential scan up to 2 V. The linear relation between the effective oxide thickness and the collision frequency extends up to around 0.8 nm. Here, the collision frequency increases by 21 meV. For thicker oxide layers the collision frequency saturates. We attribute this behavior to the completion of the interface between gold and gold oxide. Additional oxide contributes to the existing layer without changing the interface. It should be mentioned that the interface between gold and oxide still moves deeper into the gold layer with growing oxide thickness, but its interaction with electrons in the metal stays the same.

In the supporting information S7, we also apply a lossy dielectric model of the oxide to the same measurement. In this case, part of the loss can be attributed to the increasing effective oxide thickness and therefore a smaller increase in collision frequency is observed. Most importantly, the lossy oxide layer scenario leads to a decrease in collision frequency above 0.8 nm, with the maximum value of collision frequency change at 15 meV. We find this behavior implausible and conclude that the lossless dielectric more accurately represents the electrochemically grown gold oxide.

CID scales with the frequency of electron collisions with the surface. This collision frequency is proportional to the inverse of the mean free path of free electrons between two surface collisions $l_{eff}$ within the excited gold volume.[10] Therefore, we need to define the $l_{eff}$ of the system under investigation to compare to literature findings. For the excitation of gold nanoparticles, it is common to assume a constant electric field over the whole volume and calculate $l_{eff}$ based on the mean chord equation $l_{eff} = 4V/S$, where $V$ is the volume and $S$ is the surface of the structure. This formula was initially developed by Cauchy and has been applied to arbitrary shaped convex nanoparticles.[41] However, for the excitation of planar gold, the electric field and energy density within the material are not constant but exhibit exponential decay. Consequently, not all electrons are excited. To derive $l_{eff}$ in this scenario, we employ the final penetration depth $\delta$ of electric energy into the metal and approximate planar gold as a disc with thickness equal to the penetration depth and a large radius. Then the volume of excited electrons is given by $V = S\delta$, resulting in $l_{eff} = 4\delta$. For gold with $\delta_{Au} \approx 13$ nm (at 800 nm based on literature permittivity data[42]) we derive $l_{eff} = 52$ nm.

The derived $l_{eff}$ for the case of planar gold, enables us to compare our results, which show a maximum increase in collision frequency by 21 meV, against previous research. We employ the normalized parameter $A_{CID} = \Delta\Gamma_{CID} l_{eff}/v_F$[10], where $v_F$ is the Fermi velocity of free gold electrons ($v_F = 1.4 \cdot 10^6 \frac{m}{s}$[43]). This parameter describes how efficient CID changes the momentum of electrons to allow interaction with light. It is not only influenced by the nature of the interface but also its coverage. We obtain a maximum of $A_{CID}$ at 1.19 (from $\Delta\Gamma_{CID} = 21$ meV). This is significantly larger than reported from previous investigations. Foerster et al. report $A_{CID}$ between 0.33 and 0.51 from thiols on gold nanorods,[10,11] and $A_{CID}$ of 0.56, 0.34 and 0.1 from $TiO_2$, $HfO_2$ and $Al_2O_3$ atomic layer deposition coatings.[15] The presented values are calculated by us from the reported collision frequency increase and the mean free path length of the respective particles. It is an open question why the electrochemically produced gold oxide results in a much stronger CID effect than the coatings with metal-oxides or covering the surface with thiols.

Dondapati et al.[16] conducted electrochemical cycling experiments using gold nanorods in NaCl and KCl electrolytes. They measured the plasmon shift and broadening through dark-field spectroscopy and observed $A_{CID}$ between 0.59 to 0.62 in 0.1M NaCl and at 0.99 in 0.1M KCl (calculated by us from reported data in Figures 2 and 4).[16] This is a lower value than measured by us, but the electrochemically active surface area of particles might be smaller as the geometric area due to residual surfactants on the surface and the area occupied by the electric contact. Also, the use of a different (non-acid) electrolyte may explain the difference. Dondapati et al. observe that the collision frequency saturates, where the resonance frequency continues to shift with increasing potential. This corresponds to our picture of growing oxide with a stable interface. They argued that the enhanced spectral broadening (increased collision frequency) is related to the adsorption of water molecules, hydroxyl ions and dissolved anions and not oxidation. We cannot confirm this explanation for our setup and identify a direct correlation between effective oxide thickness and collision frequency, as shown in Figure 2. Additional experimental evidence that the oxidation peak causes the collision frequency increase in our experiments is shown and discussed in the supporting information S8. There, the upper limit of the potential scan was set to the region, where the main increase of collision frequency is observed. The measurements show significant hysteresis, indicating surface oxidation.

Previous spectroscopic and ellipsometric investigations of electrochemically produced gold oxide on gold[28–32] did not consider the CID effect. In contrast to our findings, they attributed the increased absorption solely to the gold oxide and therefore found it to exhibit absorbing properties. We applied the results for the complex refractive index of the gold oxide layer found by Cook and Ferguson[28] to our own ellipsometry data and found a good fit with constant collision frequency of gold. This suggests that our approach, which considers the CID effect via change in collision frequency, is not in opposition to their findings but provides a more precise way to describe the behavior at the interface.

A different type of experiment conducted by Liu et al.[36] used reactive sputtering to produce gold oxide and investigated the optical properties with spectroscopy. They observed a different optical behavior then Cook and Ferguson with the refractive index remaining almost constant at 2.75 in the visible and near-infrared spectrum, but also observe slight absorption explained by a small extinction coefficient. As discussed in the supporting information S7, we argue the gold oxide layer to be purely dielectric. We see the possibility of remaining pure gold to be the cause for the absorption they observed. Additionally, it remains uncertain if the gold oxide produced electrochemically is identical to that obtained from reactive sputtering. We employed the refractive index of 2.75 found by Liu et al. for the interpretation of our results but should mention that we cannot distinguish the refractive index and thickness of oxide layer separately in our ellipsometric fitting. Nonetheless, a change of the refractive index would only change the scale of the abscissa in Figure 3, but not change the general saturation characteristics and conclusion. A lower refractive index for the oxide would result in a fit with a thicker dielectric layer.

We observe a gold oxide layer with effective thickness between 0.6 nm and 1.2 nm from our electrochemical experiment, which is in reasonable agreement with results from Cook and Ferguson[28] and Horkans et al.[31] But it should be mentioned that Valtiner et al.[44] measured the gold oxide thickness from electrochemical oxidation (in aqueous $HNO_3$) with a direct approach and found it to be much thicker. They find layer thicknesses of approx. 3 nm and attributed the electrochemically grown layer to Au(OH)$_3$ based on thickness and charge measurements. As mentioned before, we are not able to distinguish gold oxide refractive index and thickness. Therefore, an optically less dense Au(OH)$_3$ layer of larger thickness would also fit our model.

**CONCLUSIONS**

In this work, we present a revised approach to understanding the optical properties of electrochemically oxidized gold, involving an increase in electron collision frequency of underlying gold from CID and a lossless oxide layer. With this approach, we see an expected saturation of collision frequency increase at certain effective oxide thicknesses due to stabilization of the metal-oxide interface morphology. We show that the CID effect from oxidation is stronger than previously observed from other gold surface modifications such as thiols or metal-oxides. It remains to clarify experimentally in which extent this strong CID effect is caused by temporal electron transfer and which by effective surface roughness.

The topic of the refractive index of gold oxide is still not fully resolved, but we argue that the gold oxide is purely dielectric and that absorbing properties can be explained solely from the increase in collision frequency of the underlying metal. We would like to note, that the optical effect of the dielectric coating provided by the oxide formation on gold nanostructures should be appropriately considered and not only attributed to a decreasing plasma frequency of the metal.

With our results, we show that the CID effect plays an important role in systems of oxide grown on pure gold (and in general metals). Consequently, we argue for the consideration of the CID effect in future discussions on this topic. We also observe that the use of ellipsometry for determining the optical properties of such oxides is difficult and that the ellipsometry technique should be complemented by other methods.

**Supporting Information.**

Additional experimental results and plots accompanied with brief discussion to provide further insight to the interested reader including results underlining the reversibility of the oxidation process and results obtained on sputtered gold samples (PDF).

**Acknowledgements**

This work was supported by Deutsche Forschungsgemeinschaft (DFG) through Project 192346071, SFB 986.

We disclose the use of ChatGPT 4 for the revision of the manuscript text at an early stage of the writing process.

**Competing Interests**

The authors declare no competing interests.

# Supporting Information

Table of Contents:

## S1: Fitting of ellipsometry measurements

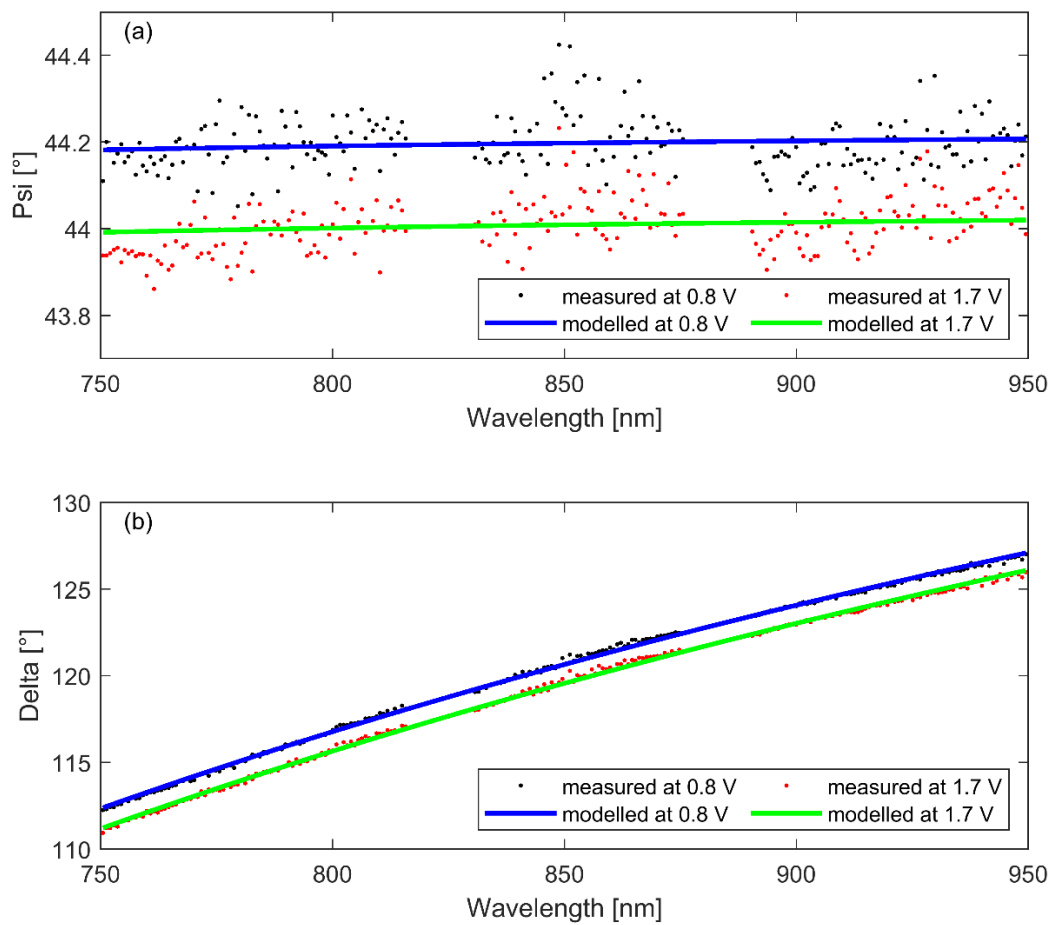

***Figure SI-1.*** *Measured ellipsometry angles Psi and Delta vs. wavelength at low and high potentials together with the fitted model values corresponding to the experiment shown in Fig. 2 of the main text. A good match between the model and the measurement is observed for both potentials.*

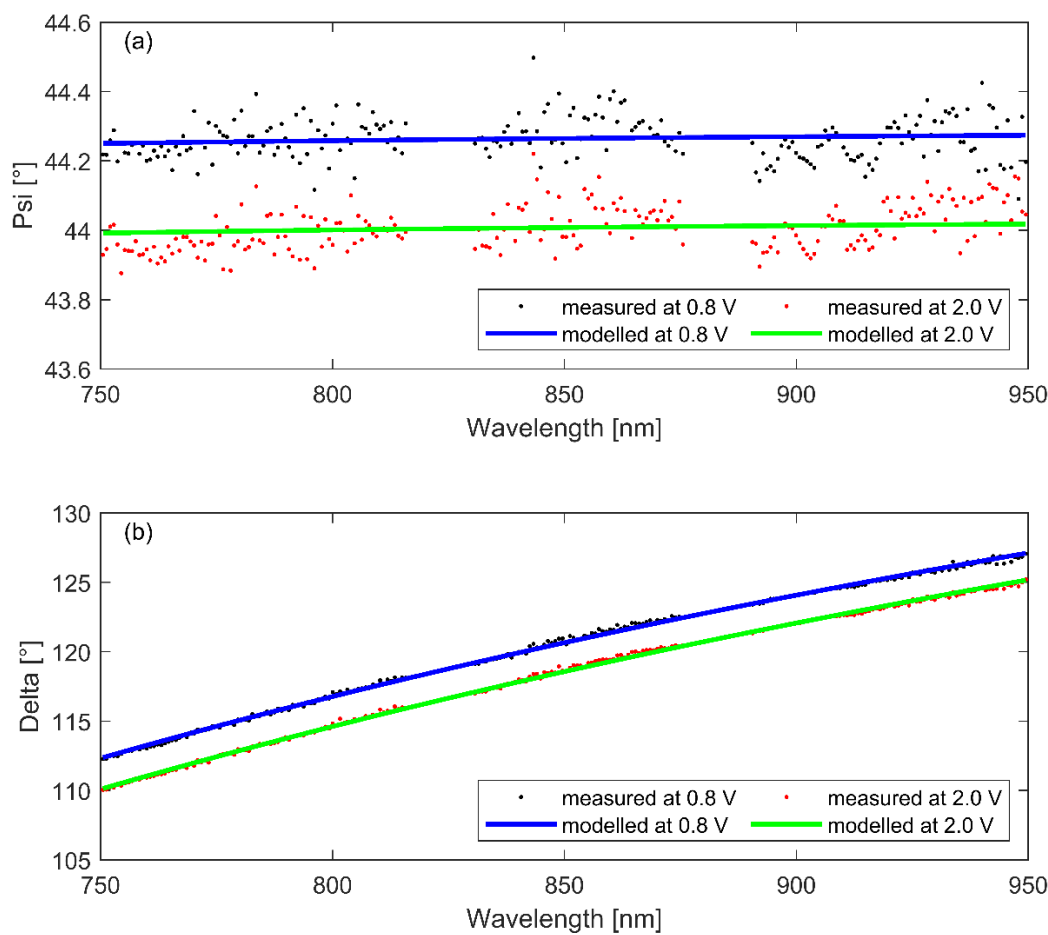

***Figure SI-2.*** *Measured ellipsometry angles Psi and Delta vs. wavelength at low and high potentials together with the fitted model values corresponding to the experiment shown in Fig. 3 of the main text. A good match between the model and the measurement is observed for both potentials.*

Fig. SI-1 and SI-2 show the measured ellipsometry angles Psi and Delta together with the best fits from our model for the experiments shown in Fig. 2 and 3 of the main text on the example of one low potential and one high potential measurement. A clear change of both Psi and Delta is observed for the gold without and with oxide. The plots show a good match between the measurement and the fitted model. The oxidation results in a decline of Psi and Delta.

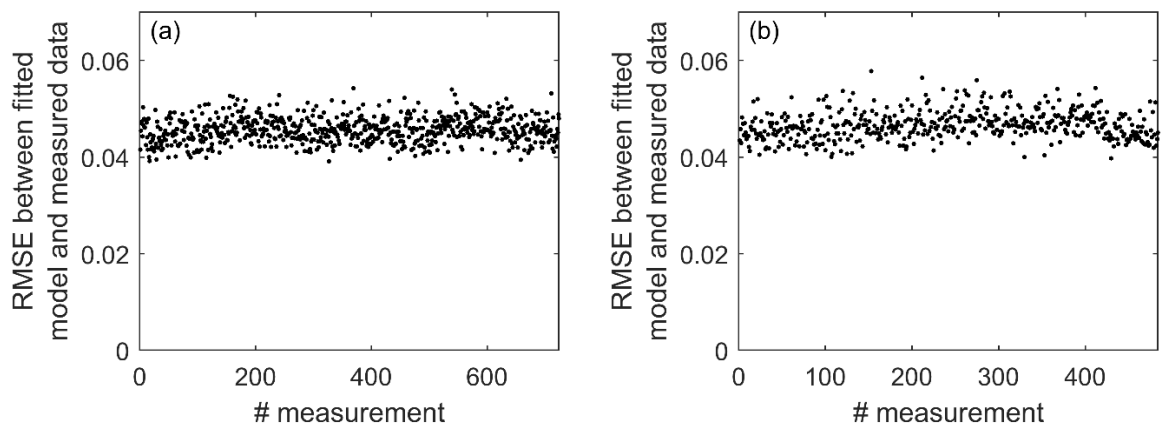

***Figure SI-3.*** *RMSE between the measured ellipsometry angles and the modelled ones. The left plot a) shows the RMSE values for all fits obtained from the experiment depicted in Fig. 2 of the main text, while b) shows those corresponding to Fig. 3 in the main text.*

Fig. SI-3 shows RMSE between the measured data and the modelled over the whole oxidation and reduction cycle. Together with Fig. SI-1 and SI-2 it shows the validity of our model over all the obtained measurements at all potentials.

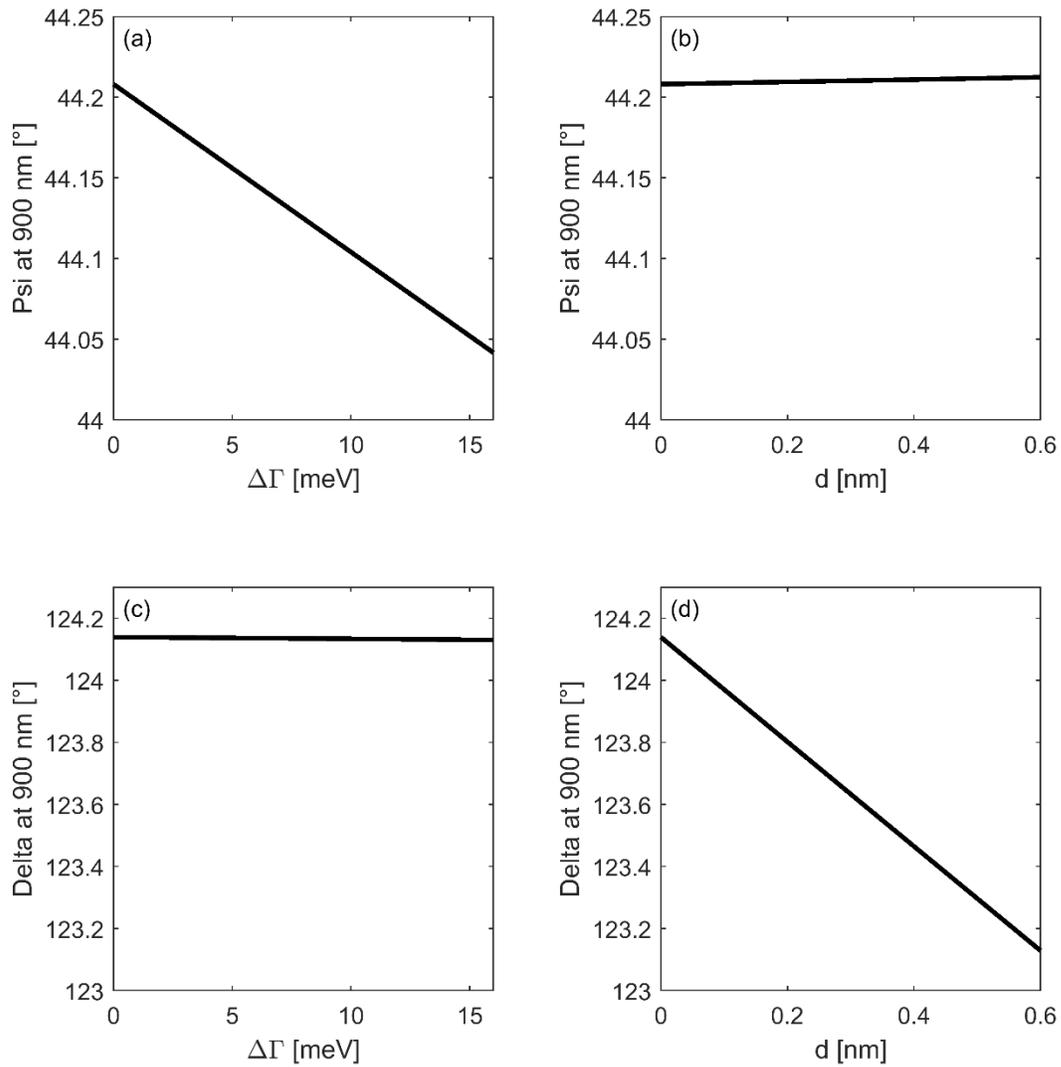

***Figure SI-4.*** *Modelled Psi and Delta values at 900 nm for the range of the collision frequency $\Gamma$ and the effective oxide thickness d of the experiment shown in Fig. 2 of the main text.*

Fig. SI-4 shows how Psi and Delta depend on the two model parameters of collision frequency and effective oxide thickness. a) and b) give insight on how Psi is influenced by a change of collision frequency $\Gamma$ and a change of oxide thickness *d*. It shows that Psi is mainly influenced by the collision frequency $\Gamma$. c) and d) give insight on how Delta is influenced by a change of collision frequency $\Gamma$ and a change of oxide thickness d. It shows that Delta is mainly influenced by the oxide thickness. This shows little cross correlation between the two fitting parameters.

## S2: Correction of the electrochemical current

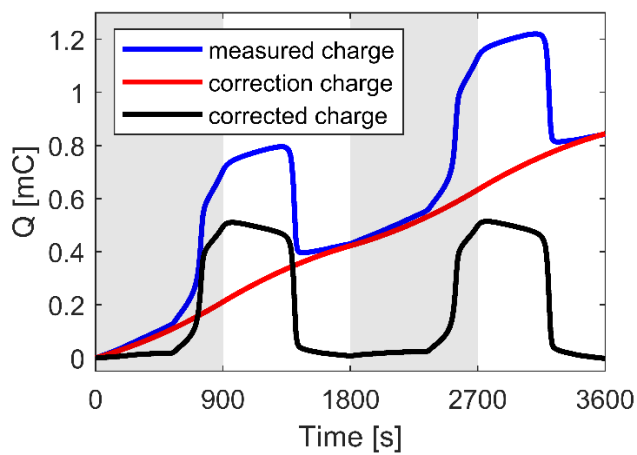

*Figure SI-5. Measured (blue), correction (red) and corrected (black) charge corresponding to the experiment shown in Fig. 2 of the main text over the span of two cyclic voltammetry scans. The grey shaded area indicates the anodic scan region and the white area the cathodic scan region.*

The measured current and thus also the measured charge comprises of the net charge $Q$ that is related to the oxidation process and a charge that originates from a background current $I_b$. We assume this current to be governed by a constant resistance $R$: $I_b = \frac{U(t)}{R}$. From this current one can obtain the correction charge $Q_c(t) = \int_{t_0}^{t} \frac{U(t)}{R} dt = \int_{t_0}^{t} \frac{U_0 + \Delta U \cdot t}{R} dt$ with the starting potential $U_0$ and the scan rate $\Delta U$. The net charge $Q$ is then obtained from: $Q(t) = Q_m(t) - Q_c(t)$. The constant resistance is obtained based on the condition that the net charge $Q$ related to oxidation is to return to zero following a complete cycle. Fig. SI-5 shows the three different charges as example from the experiment depicted in Fig. 2 of the main text. The slight waviness seen in the correction charge curve results from the periodical increase and decrease of the applied potential which modulates the accumulated charge.

## S3: Reversibility of the CID effect from oxide growth over 11 potential cycles

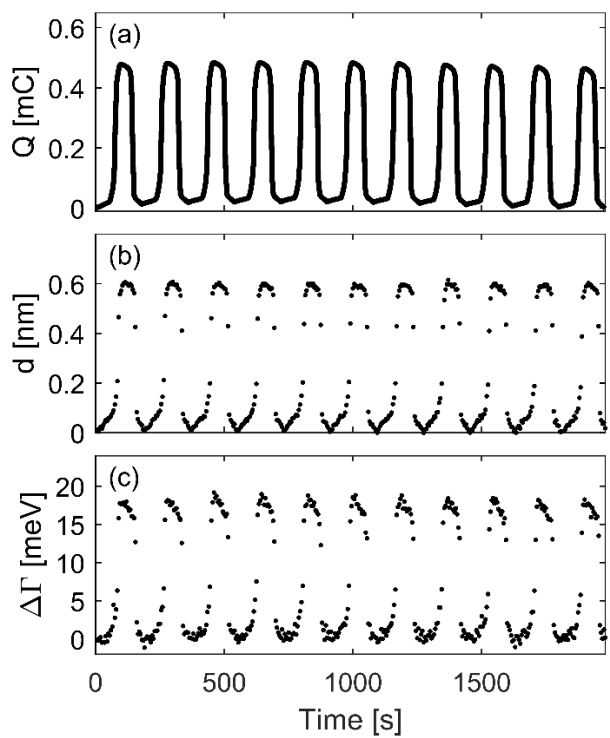

*Figure SI-6.* *Reversible variations/change in (a) net charge Q, (b) oxide thickness d and (c) collision frequency of template stripped Au(111) Γ during 11 CV scans between 0.8 and 1.7 V vs. RHE at a scan rate of 10 mV/s in 0.5M $H_2SO_4$. The top subfigure (a) shows the net charge corrected by the background current. The middle (b) and bottom (c) plot show the fitting results from the ellipsometry measurements, displaying the variation of the oxide thickness and the change of collision frequency, respectively. They indicate a well reversible oxidation process along with a reversible change of the optical properties of the Au layer from CID. The electrochemical oxidation results in a 0.6 nm thick oxide layer and an increase in collision frequency of 18 meV.*

Fig. SI-6 shows results of the ellipsometry measurements obtained during potential cycling with 10 mV/s scan rate for 11 cycles from the first experiment discussed in the main text. The top subfigure (a) shows the oxidation and reduction from electrochemical cycling, while the middle (b) and bottom (c) ones show the fitting results from ellipsometry. These reveal a reversible oxidation process and show correlation of the corrected net charge with the oxide layer thickness as well as the change of collision frequency. The increase in collision frequency shows CID from the interface of the electrochemical produced oxide layer and the gold surface. The oxide layer during this experiment has a maximum thickness of 0.6 nm and the increase in collision frequency reaches values up to 18 meV. In addition to the experiment shown in Fig. 2 of the main text, this one extends over 11 cycles to underline the reversibility of the oxide growth and the collision frequency change.

S4: Linear relation between change in collision frequency and oxide thickness

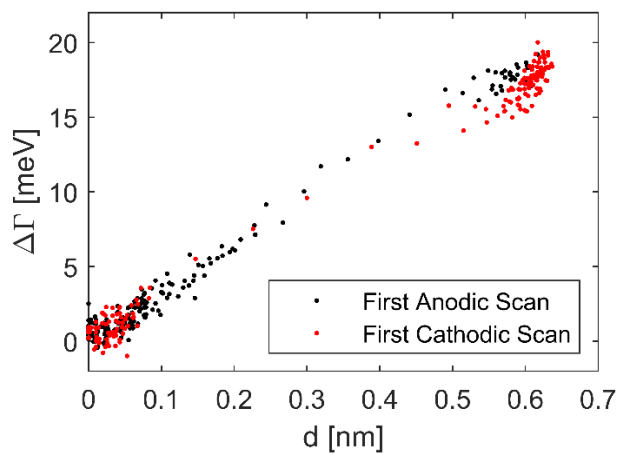

*Figure SI-7.* *Collision frequency change of template stripped Au(111) ΔΓ vs. oxide thickness d during the first CV scan (from Fig. 2 in main text) between 0.8 and 1.7 V vs. RHE in 0.5M $H_2SO_4$. The black dots indicate the anodic (positive) scan direction and the red ones the cathodic (negative) scan direction. This plot reveals the linear relation between the two parameters in both scan directions.*

Fig. SI-7 shows the linear relation between oxide thickness and change of collision frequency for the second experiment depicted in Fig. 2 of the main text. A similar linear relationship was observed with CID from thiols by Foerster et al.[1] The reversibility of the process is clearly seen in the cathodic scan. In the cathodic scan direction, a small dip is observed, which corresponds to the slight decline of collision frequency before the reduction mentioned in the main text.

## S5: Additional plots of collision frequency and oxide thickness vs. charge and potential

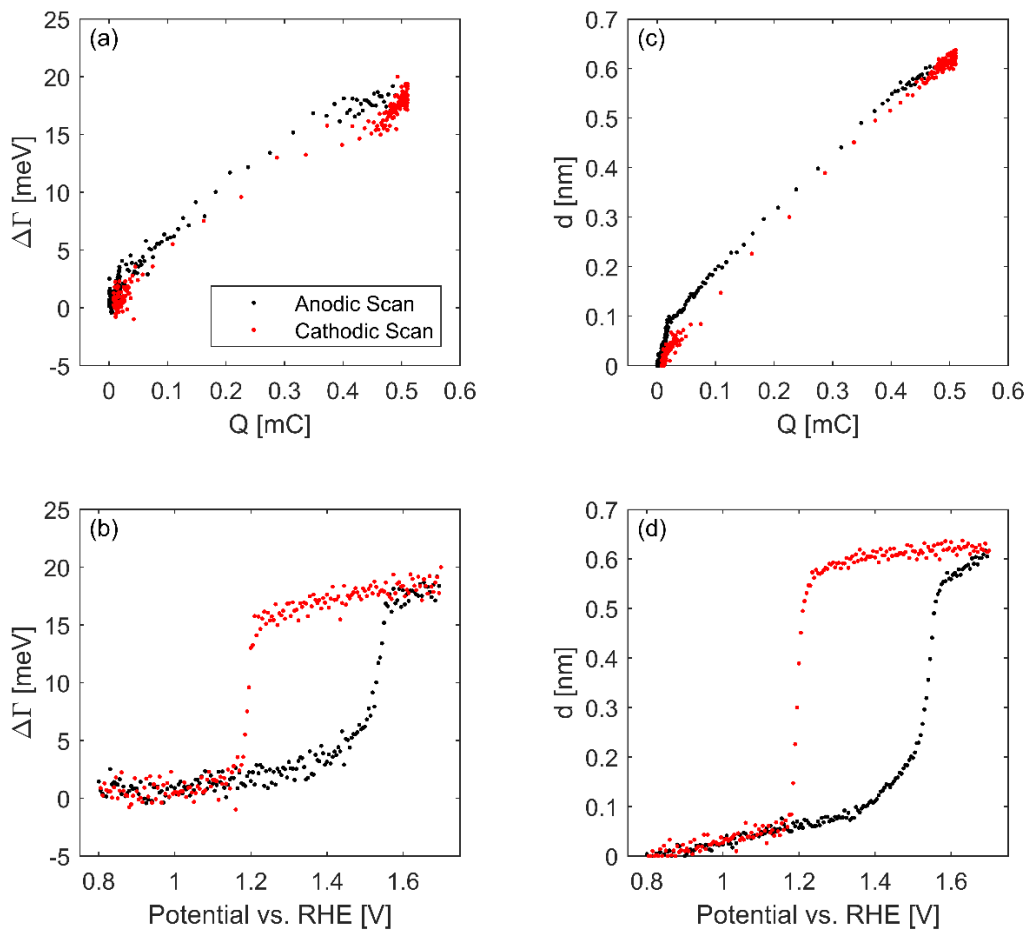

*Figure SI-8.* *Plots of data on template stripped Au(111) obtained during the first CV scan with a 1 mV/s scan rate in 0.5M $H_2SO_4$ showing change in collision frequency $\Delta\Gamma$ vs. net charge Q (a) and vs. potential (b) as well as oxide thickness d vs. net charge Q (c) and vs. potential (d). The black dots indicate the anodic (positive) scan direction and the red ones the cathodic (negative) scan direction.*

Fig. SI-8 shows the ellipsometric fitting results (change in collision frequency $\Delta\Gamma$ and oxide thickness *d*) from the first CV scan depicted in Fig. 2 of the main text vs. the net charge *Q* and the potential. The plots vs. charge reveal a steep rise at close to zero net charge for both the collision frequency and the oxide layer thickness followed by a linear behavior.

We argue that the steep rise originates from adsorption of oxygen species below oxidation potentials[2] and that the related current is masked by our correction for the background current. The plots vs. potential show this behavior up to approx. 1.4 V. We observe a steady rise of both fitted parameters in this region. At higher potentials, the oxidation occurs and reveals the distinct hysteresis between the oxidation and reduction potentials.

S6: Cyclic Voltammogram of experiment depicted in Fig. 3 of the main text

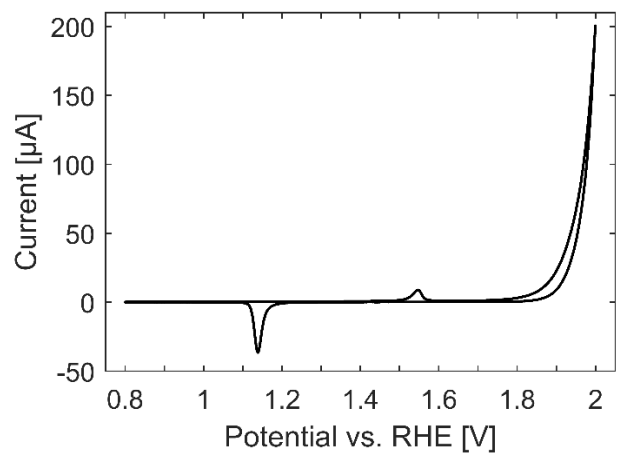

*Figure SI-9.* *Cyclic voltammogram of template stripped Au(111) in 0.5M $H_2SO_4$ with a scan rate of 1 mV/s between 0.8 and 2 V vs. RHE corresponding to the results depicted in Fig. 3 of the main text. The steep rise above 1.8 V corresponds to the onset of the oxygen evolution reaction.*

Fig. SI-9 shows the cyclic voltammogram corresponding to the third electrochemical experiment depicted in Fig. 3 of the main text. With this experiment, we aim to obtain thicker oxide layers by applying a higher upper vertex potential of 2 V. Above 1.8 V vs. RHE we observe a steep rise of current that corresponds to the oxygen evolution reaction.

S7: Comparison of fitting results for a purely dielectric and a lossy dielectric oxide layer in anodic and cathodic scan direction

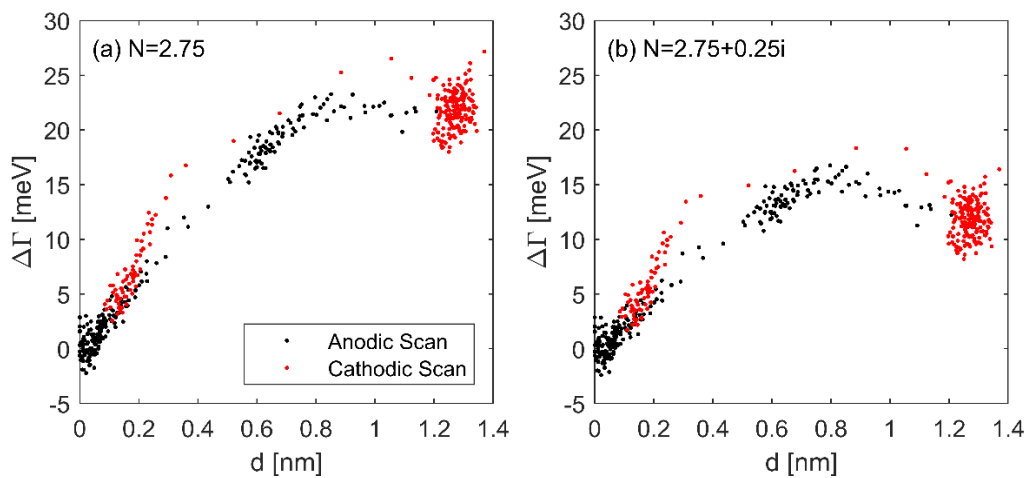

*Figure SI-10.* Comparison of the fitting results for a purely dielectric (a) and a lossy dielectric gold oxide (b) for both anodic (black) and cathodic (red) scan direction of template stripped Au(111) in 0.5M $H_2SO_4$. We show the change of the collision frequency $\Delta\Gamma$ vs. the oxide thickness d. For the lossy dielectric gold oxide, the collision frequency does not only saturate but also declines at larger thicknesses. The fitting results for the cathodic scan reveal a nonreversible process as both the collision frequency and the oxide thickness do not reach their initial values.

In the main text discussion, we consider a purely dielectric gold oxide from electrochemical oxidation and use the value of $n = 2.75$ found by Liu et al. omitting the small extinction coefficient they report.[3] They used reactive sputtering for their study. Fig. SI-10 shows that a lossy dielectric for the gold oxide (with $k = 0.25$ as reported by Lui et al.[3]) results in a decline of the collision frequency instead of a saturation, which we observe for the purely dielectric oxide assumed in the main text. Considering our underlying model and the understanding of CID, we see this as an unphysical behavior and argue that the gold oxide is indeed purely dielectric. We argue that some not oxidized metallic gold may have caused the observed losses in the study of Liu et al. Another study on the reactive sputtering of gold in a pure oxygen atmosphere by Pierson et al.[4] found the technique to not be applicable as they observed deposition of pure gold. We need to mention that when employing the lossy model for gold oxide, we observe an overall smaller change of the collision frequency from CID.

For this experiment, the fitting results from the cathodic scan do not match the ones from the anodic one. Both the change of collision frequency and the oxide thickness do not return to zero, which indicates a nonreversible process due to high potentials in the oxygen evolution regime. We suspect that residual oxygen trapped in the gold and/or induced residual surface roughness are responsible for the offset.

## S8: Additional experiments indicating oxidation as cause for CID in our setup

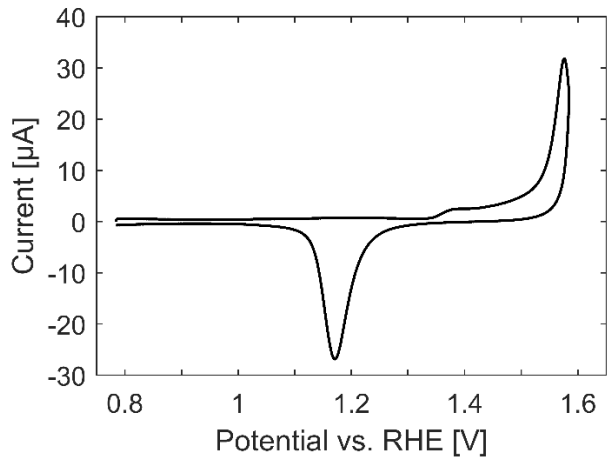

*Figure SI-11.* *Cyclic voltammogram of template stripped Au(111) in 0.5M $H_2SO_4$ with a scan rate of 10 mV/s between 0.785 and 1.585 V vs. RHE. The upper scan limit was set into the middle of the oxidation process.*

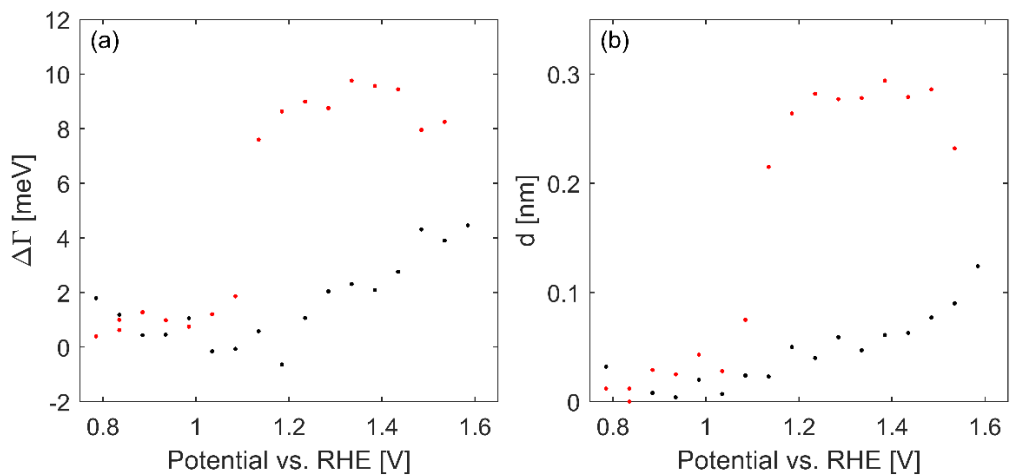

*Figure SI-12.* *Fitted change in the collision frequency ΔΓ (a) and the oxide thickness d (b) vs. potential for the CV scan shown in Fig. SI-11. The black dots indicate the anodic and the red ones the cathodic scan direction. Note that at the beginning of the cathodic scan both collision frequency and thickness continue to increase as the scanning back to lower potentials still results in a positive current for a short amount of time. This can also be seen on the right side of Fig. SI-11.*

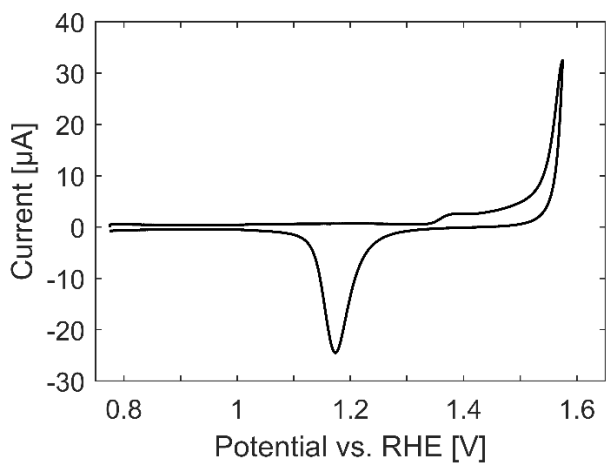

*Figure SI-13.* Cyclic voltammogram of template stripped Au(111) in 0.5M $H_2SO_4$ with a scan rate of 10 mV/s between 0.775 and 1.575 V vs. RHE. The upper scan limit was set into the middle of the oxidation process.

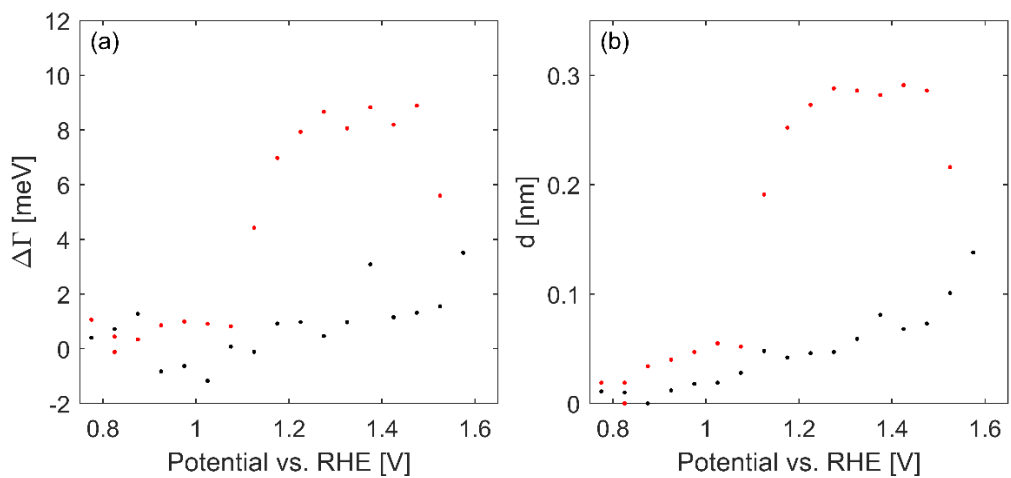

*Figure SI-14.* Fitted change in the collision frequency $\Delta\Gamma$ (a) and the oxide thickness d (b) vs. potential for the CV scan shown in Fig. SI-13. The black dots indicate the anodic and the red ones the cathodic scan direction. Note that at the beginning of the cathodic scan both collision frequency and thickness continue to increase as the scanning back to lower potentials still results in a positive current for a short amount of time. This can be seen on the right side of Fig. SI-13.

Dondapati et al.[5] find that the resonance broadening of electrochemically oxidized gold nanorods (and corresponding increase of collision frequency) is not related to the oxide formation for their experiments in 0.1M NaCl and 0.1M KCl. They come to this conclusion as they don't observe a significant hysteresis of broadening when only applying potentials where they observe the main broadening (Fig. 3b from Ref. [5]). We conducted a similar experiment with our setup, by conducting CV scans together with ellipsometry measurements with the upper potential limit set to 1.585 and 1.575 V vs. RHE. The results are shown in Fig. SI-11 to SI-14. Both upper limits are set within the region where we observe the main increase in the collision frequency. For both experiments, we observe a hysteresis of the collision frequency as well as the oxide thickness. Both decline significantly when reaching the reduction potential in the cathodic scan. With these experiments, we only achieve an increase in collision frequency of approx. 9 meV and thus are in the middle of the regime where the interface between gold and gold oxide changes and CID occurs. Consequently, we find that for our experiments it is the oxide layer that is responsible for the increase in collision frequency.

## S9: Experimental results obtained from sputtered polycrystalline Au

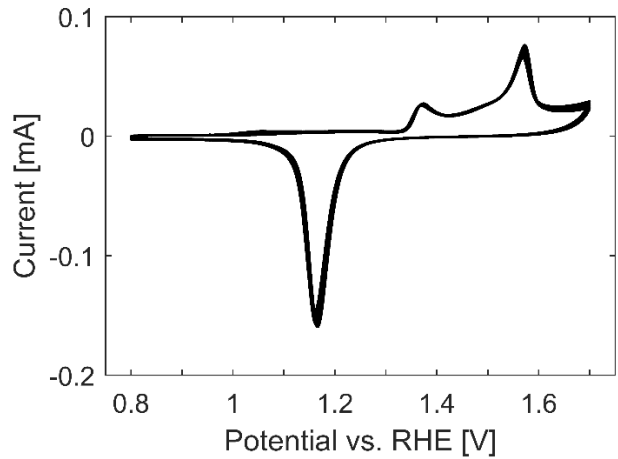

*Figure SI-15.* *Cyclic voltammogram of sputtered polycrystalline gold in 0.5M $H_2SO_4$ with a scan rate of 10 mV/s and 10 scans between 0.8 and 1.7 V vs. RHE.*

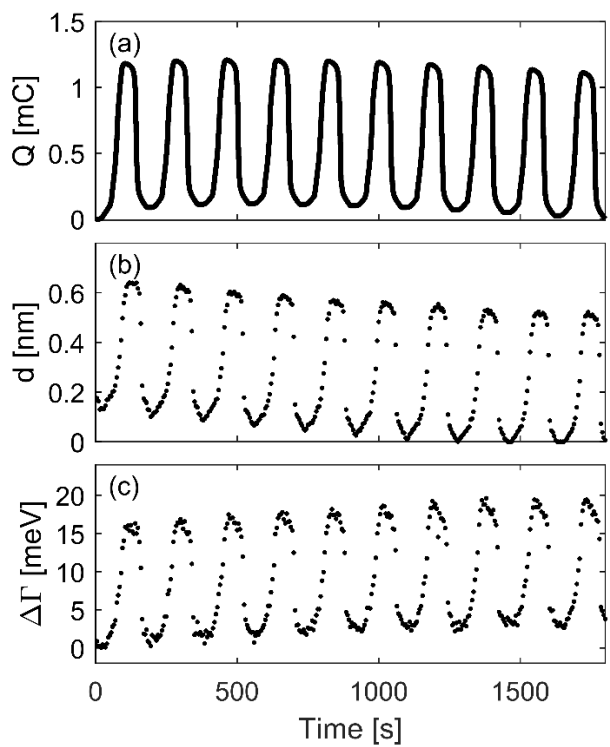

*Figure SI-16.* *Reversible variations/change in (a) net charge Q, (b) oxide thickness d and (c) collision frequency of sputtered polycrystalline Au $\Delta\Gamma$ during 10 CV scans between 0.8 and 1.7 V at a scan rate of 10 mV/s in 0.5M $H_2SO_4$. The top subfigure (a) shows the net charge corrected by the background current. The middle (b) and bottom (c) plots show*

*the fitting results from the ellipsometry measurements, displaying the variation of the oxide thickness and the change in the collision frequency, respectively. Interestingly, we need to set the initial oxide thickness to 0.15 nm and observe an overall decline in the oxide thickness and a rise in the collision frequency over multiple cycles.*

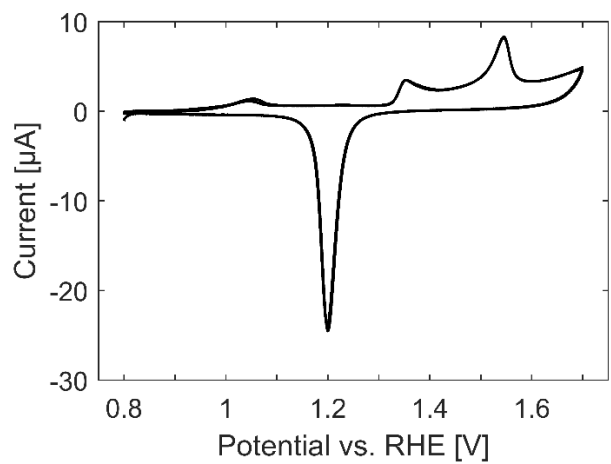

*Figure SI-17. Cyclic voltammogram of sputtered polycrystalline gold in 0.5M $H_2SO_4$ with a scan rate of 1 mV/s and 2 scans between 0.8 and 1.7 V vs. RHE.*

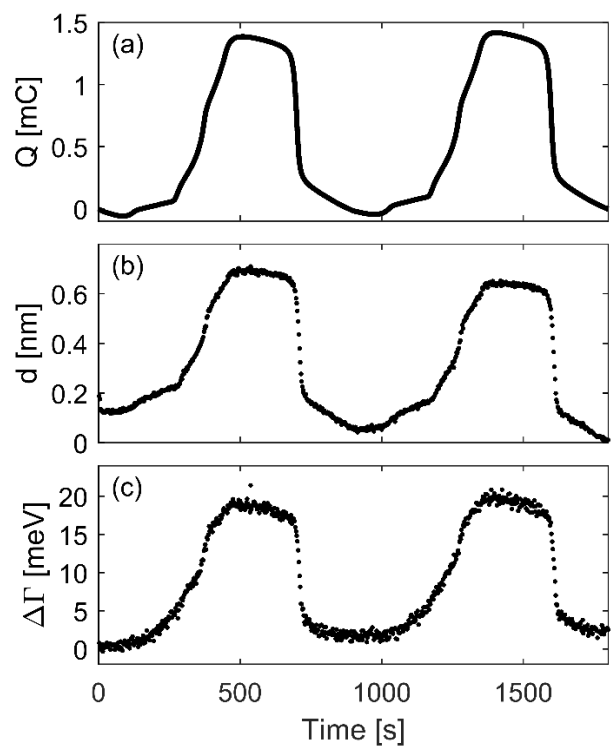

***Figure SI-18.*** *Reversible variations/change in (a) net charge Q, (b) oxide thickness d and (c) collision frequency of sputtered polycrystalline Au ΔΓ during 2 CV scans between 0.8 and 1.7 V vs. RHE at a scan rate of 1 mV/s in 0.5M $H_2SO_4$. The top subfigure (a) shows the net charge corrected by the background current. The middle (b) and bottom (c) plots show the fitting results from the ellipsometry measurements, displaying the variation of the oxide thickness and the change in the collision frequency respectively. Interestingly we need to set the initial oxide thickness to 0.15 nm and observe an overall decline in oxide thickness and a rise in the collision frequency over multiple cycles. The electrochemical oxidation results in the reversible adsorption and desorption of a 0.6 nm-thick oxide layer and an increase in the collision frequency of 18 meV.*

For sputtered polycrystalline gold (100 nm Au on Ti adhesion layer on Si-wafer substrate) we observe similar CID effect compared to results for template-stripped Au(111) discussed in the main text. The CV scans (Fig. SI-15 and SI-17) reveal the existence of different crystal orientations.[6] Fig. SI-16 and SI-18 show the variations in the oxide thickness and the collision frequency along with the net charge. The total increase in collision frequency is approx. the same as for Au(111), while the overall change in oxide thickness during one cycle is slightly smaller. Interestingly we had to set the initial value for the oxide to 0.15 nm as the overall oxide thickness declines from potential cycling. Also, we observe a slight increase in collision frequency by the cycling, but the change within a cycle remains approx. the same.